\documentclass[openany]{svmult}

\usepackage{makeidx}         
\usepackage{graphicx}        
\usepackage{multicol}        
\usepackage[bottom]{footmisc}
\usepackage{cite}
\usepackage{amsmath,amssymb}
\usepackage{pifont}
\usepackage{enumerate}

\begin{document}

\title*{A Brief History of Economics: An Outsider's Account}
\author{Bikas K Chakrabarti}
\institute{Theoretical Condensed Matter Physics Division and
Centre for Applied Mathematics and Computational Science,
Saha Institute of Nuclear Physics,
1/AF, Bidhannagar, Kolkata 700 064, India.\\
\texttt{bikask.chakrabarti@saha.ac.in}}

\maketitle

\vskip 1 cm

\abstract{A dangerously brief history of the developments of
the main ideas in economics, as observed by a physicist, is given. 
This was published in \textit{Econophysics of Stock and
Other Markets}, Eds. A. Chatterjee, B. K. Chakrabarti, 
New Economic Windows Series, Springer, Milan, 2006, pp~219-224.}

\vskip 1.0cm
\noindent
When physics started to develop, say with Galileo Galelei (1564-1642),
there were hardly any science at a grown-up stage to get help or
inspiration from. The only science that was somewhat
grown up was mathematics, which is an analytical science (based on logic)
and not synthetic (based on observations/ experiments carried out in
controlled environments or laboratories). Yet, developments in mathematics,
astronomical studies in particular, had a deep impact in the development
of physics, of which the (classical) foundation was almost single-handedly
laid down by Isaac Newton (1643-1727) in the seventeenth and early eighteenth
century. Mathematics remained at the core of physics since then.
The rest of ``main stream'' sciences, like chemistry, biology etc all tried
to get inspiration from, utilize, and compare with physics since then.

To my mind, development in social sciences started much later.
Even the earliest attempt to model an agricultural economy in a kingdom,
the ``physiocrats' model", named after the profession of its pioneer, 
the french royal physician Francois
Quesnay (1694-1774), came in the third quarter of the eighteenth century
when physics was already put on firm ground by Newton. The physiocrats
made the observation that an economy consists of the components like
land and farmers, which are obvious. Additionally, they identified the
other components as investment (in the form of seeds from
previous savings) and protection (during harvest and collection, by the 
landlord or the king). The impact of the physical sciences, in emphasizing
these observations regarding components of an economy, is clear.
The analogy with human physiology then suggested that, like the healthy 
function of a body requiring proper functioning of each of its components
or organs and the (blood) flow among them remaining uninterrrupted,
each component of the economy should be given proper care (suggesting
rent for land and tax for protection!). Although the physiocrats'
observations were appreciated later, the attempt to conclude using the
analogy with human physiology was not.

Soon, at their last phase, Mercantilists, like Wilhelm von Hornick (1638-1712),
James Stewart (1712-1780) et al, made some of the most profound and emphatic
observations in economics, leading to the foundation of political economy.
In particular, the observations by the British merchants (who traded in the
colonies, including India, in their own set terms) that 
instability/unemployment growing at their home country in years whenever there
had been a net trade deficit and out-flow of gold (export being less than 
import). This led to the formulation of the problem of effective demand:
even though the merchants, or traders were independently trading (exporting or
importing goods) with success, the country's economy as a whole did not do
well due to lack of overall demand
when there was a net flow of gold (the international exchange medium)
to balance the trade deficit! This remains still a major problem in
macroeconomics. The only solution in those days was to introduce tax on import:
the third party (namely the government) intervention on individuals' choice
of economic activity (trade). This immediately justified the involvement of
the government in the economic activities of the individuals.

In a somewhat isolated but powerful observation, Thomas Malthus (1766-1834) 
made a very precise modelling of the conflict between agricultural production
and population growth. He assumed that the agricultural production can
only grow (linearly) with the area of the cultivated land. With time $t$,
say year, the area can only grow linearly ($\propto t$) or in arithmetic
progression (AP). The consumption depends on the population which,
on the other hand, grows exponentially ($\exp [t]$) or in geometric
progression (GP). Hence, with time, or year $1,2,3,\ldots$, the agricultural
production grows as $1,2,3,\ldots$, while the consumption
demand or population grows in a series like $2,4,8,\ldots$. No matter, how
much large area of cultivable land we start with, the population
GP series soon takes over the food production AP series and the 
population faces a disaster --- to be settled with
famine, war or revolution! They are inevitable, as an exponentially
growing function will always win over a lineraly growing function
and such disasters will appear almost periodically in time!

Adam Smith (1723-1790) made the first attempt to formulate the economic science.
He painstakingly argued that a truely many-body system of selfish agents, 
each having no idea of benevolence or charity towards its fellow neighbours, or 
having no foresight (views very local in space and time), can indeed 
reach an equilibrium where the economy as a whole
is most efficient; leading to the best acceptable price for each commodity.
This `invisible hand' mechanism of the market to evolve towards the 
`most efficient' (beneficial to \textit{all} participating agents) predates
by ages the demonstration of `self-organisation' mechanism in physics
or chemistry of many-body systems, where each constitutent cell or
automata  follows very local (in space and time) dynamical rules
and yet the collective system evolves towards a globally `organised' pattern 
(cf. Ilya Prigogine (1917-), Per Bak (1947-2002) et al). This idea of
`self-organizing or self-correcting economy' by Smith of course
contradicted the prescription of the Mercantilists regarding government
intervention in the economic activities of the individuals, and argued
tampering by any external agency to be counterproductive. 

Soon, the problem of price or value of any commodity in the market became 
a central problem. Following David Ricardo's (1772-1823) formulation of
rent and labour theory of value, where the price depends only on the
amount of labour put by the farmers or labourers, Karl Marx (1818-1883)
formulated and forwarded emphatically the surplus labour 
theory of value or wealth in any economy. However, none of them could solve 
the price paradox: why diamond is costly, while coal is cheap? The amount of
labour in mining etc are more or less the same for both. Yet, the prices
are different by astronomical factors! This clearly demonstrates the failure
of the labour theory of value. The 
alternative forwarded was the utility theory of price: the more the utility
of a commodity, the more will be its price.
But then, how come a bottle of water costs less than a bottle of wine?
Water is life and certainly has more utility! 
The solution identified was marginal utility.
According to marginal utility theory, not the utility but rather its
derivative with respect to the quantity determines the price: water is cheaper
as its marginal utility at the present level of its availability is less than
that for wine --- will surely change in a desert.
This still does not solve the problem completely. Of course increasing
marginal utility creates increasing demand for it, but its price must 
depend on its supply (and will be determined by equating the demand 
with the supply)!
If the offered (hypothetical) price $p$ of a commodity increases, the supply
will increase  and the demand for that
commodity will decrease. The price, for which
supply $S$ will be equal to demand $D$, will be the market price
of the commodity: $S(p)=D(p)$ at the market (clearing) price.
However, there are problems still.   Which demand should be equated
to which supply? It is not uncommon to see often (in India) that price as well
as the demand for rice (say) increases simultaneously. This can occur when
the price of the other staple alternative (wheat) increases even more.

The solutions to these problems led ultimately to the formal development of
economic science in the early twentieth century
by L\'eon Walras (1834-1910), Alfred Marshal (1842-1924) and others:
marginal utility theory of price and cooperative or coupled (in all
commodities) demand and supply equations.
These formulations went back to the self-organising picture of
any market, as suggested by Adam Smith, and incorporated this marginal
utility concept, and utilized these coupled demand-supply equations:
$D_i (p_1,p_2,\ldots,p_i,\ldots,p_N,M) = S_i (p_1,p_2,\ldots,p_i,\ldots,p_N,M)$
for $N$ commodities and total money $M$ in the market, each having relative
price tags $p_i$ (determined by marginal utility rankings) 
and demand $D_i$ and supply $S_i$; $i=1,2,\ldots,N$ and the
functions $D$ or $S$ are in general nonlinear in their arguments.
These formal and abstract formulations of economic science were not 
appreciated very much in its early days and had a temporary setback.
The lack of acceptance was due to the fact that neither utility nor marginal
utility is measurable and the formal solutions of these coupled
nonlinear equations in many ($p_i$) variables still remain elusive.
The major reason for the lack of appreciation for these formal theories was
a profound and intuitive obsevation by John Maynard Keynes (1883-1946)
on the fall of aggregate (or macroeconomic) effective demand in the market
(as pointed out earlier by the Mercantilists; this time due to `liquidity
preference' of money by the market participants) during the great depression of
1930's. His prescription was for government intervention (in direct 
contradiction with the `laissez-faire' ideas of leaving the market to its
own forces to bring back the equilibrium, as Smith, Walras et al proposed)
to boost aggregate demand by fiscal measures. This prescription made
immediate success in most cases. By the third quarter of the twentieth 
century, however, its failures bacame apparent and the formal developments in
microeconomics took the front seat again.

Several important, but isolated observations in the meantime contributed
later very significantly. Vilfredo Pareto (1848-1923) observed that the number
density $P(m)$ of riches in any society decreases rather slowly with their
richness $m$ (measured in those days by palace sizes, number of horses, 
etc of the kings/landlords in all over Europe): $P(m) \sim m^{-\alpha}$;
for very large $m$ (very rich people); $2 < \alpha < 3$ (Cours d'Economic
Politique, Lausanne, 1897). It may be mentioned, at almost the same time,
Joshiah Willard Gibbs (1839-1903) had put forward precisely that the number
density $P(\epsilon)$ of particles (or microstates) with energy 
$\epsilon$ in a thermodynamic
ensemble in equilibrium at temperature $T$ falls off much faster: 
$P(\epsilon) \sim \exp[-\epsilon/T]$ 
(Elementary Principles of Statistical Mechanics, 1902).
This was by then rigorously established in physics. The other important
observation was by Louis Bachelier (1870-1946) who modelled the speculative
price fluctuations ($\sigma$), over time $\tau$, using a Gaussian statistics 
(for random walk): $P(\sigma) \sim \exp[-\sigma^2/\tau]$
(Thesis: Th\'eorie de la Sp\'eculation, Paris, 1900). This actually
predated Albert Einstein's (1879-1955) random walk theory (1905) by five years.
In another isolated development, mathematician John von Neumann (1903-1957)
started developing the game theories for microeconomic behavior of
partners in oligopolistic competitions (to take care of the strategy 
changes by agents, based on earlier performance).

In the mainstrem economics, Paul Samuelson (1915-) investigated the dynamic
stabilities of demand-supply equilibrium by formulating, following
Newton's equations of motion in mechanics, dynamical equations 
$\frac{d D_i}{d t}=\sum_i J_{ij} D_j (p_1,p_2,\ldots,p_N,M)$ and
$\frac{d S_i}{d t}=\sum_i K_{ij} S_i (p_1,p_2,\ldots,p_N,M)$,
with the demand and supply (overlap) matrices $\underline{J}$ and 
$\underline{K}$ respectively for $N$ commodities, and by looking for
the equilibrium state(s) where $dS/dt =0= dD/dt$
at the market clearing prices $\{p\}$ where $D_i(\{p\},M)=S_i(\{p\},M)$. 
Jan Tinbergen (1903-1994),
a statistical physicist (student of Paul Ehrenfest of Leiden University)
analysed the business cycle statistics and initiated the formulation
of econometrics. By this time, these formal developments in economics, with
clear impact of other developed sciences (physics in particular), were
getting recognized. In fact, Tinbergen was the first recipient of the newly
instituted Nobel prize in Economics in 1969 (for other sciences, they
started in 1901; a delay by 68 years in 105 years' history of the prize!) 
and the next year, the prize went
to Samuelson. Soon, the formal developments like the axiomatic foundations
of utility (ranking) theory, and solution of general equilibrium theory
by Kenneth Arrow (1921-), those of George Stigler (1911-1991), who first
performed Monte Carlo simulations of markets (similar to those of
thermodynamic systems in physics), or that of John Nash (1928-), giving the
proof of the existence of equilibrium solutions in strategic games, etc,
all were appreciated by awarding the Nobel prizes in economics
(in 1972, 1982 and 1994 respectively). Although the impact of
developments in physics had a clear mark in those of economics so far, it
was not that explicit until about a decade and a half back.

The latest developments (leading to econophysics) 
had of course its seed in several 
earlier observations. Important among them was by Benoit Mandelbrot (1924-)
when he observed in 1963 that the speculative fluctuations (in the cotton 
market for example) have a much slower rate of decay, compared to that
suggested by the Gaussian statistics of Bachelier, and falls down following 
a power
law statistics: $P(\sigma) \sim \sigma^{-\alpha}$ with some robust exponent
value ($\alpha$) depending on the time scale of observations. With the 
enormous amount  of stock market data now available on the internet,
Eugene Stanley, Rosario Mantegna and coworkers established firmly
the above mentioned (power-law) form of the stock price fluctuation 
statistics in late 1990's. Simultaneously, two important modelling efforts,
inspired directly from physics, started: the minority game models, for
taking care of contigious behavior (in contrast to perfect rational behavior)
of agents in the market, and learning from the past performance of the
strategies, were developed by Brian Arthur, Damien Challet, Yi-Cheng Zhang
et al, starting 1994. The other modelling effort was to capture the income 
or wealth distribution in society, similar to energy distributions in
(ideal) gases. These models intend to capture both the initial 
Gamma/log-normal distribution for the income distributions of poor and 
middle-income groups and also the Pareto tail of the distribution for the 
riches. It turned out, as shown by the Kolkata group during 
the last half of 1990
to the first half of 2000, a random saving gas model can easily capture these
features of the distribution function. However, the model had several
well documented previous, somewhat incomplete, versions available for a long
time. Meghnad Saha (1893-1956), the founder of Saha Institute of
Nuclear Physics, Kolkata (named so after its founder's death), 
and collaborators, already discussed at length in their text book,
in the 1950's, the possibility of using Maxwell-Boltzmann velocity 
distribution (a Gamma distribution) in an ideal gas to represent the income
distribution in societies: ``suppose in a country, the assessing department
is required to find out the average income per head of the population. 
They will proceed somewhat in the similar way ... (the income distribution)
curve will have this shape because the number of absolute beggers is
very small, and the number of millionaires is also small, while the majority
of the population have average income.'' (section on `Distribution of
velocities' in A Treatise on Heat, M. N. Saha and B. N. Srivastava, 
Indian Press, Allahabad, 1950; pp. 132-134). This modelling had the obvious
drawback that the distribution could not capture the Pareto tail.
However, the accuracy of this Gibbs distribution for fitting the income
data available now in the internet has been pointed out recently by Victor 
Yakovenko and collaborators in a series of papers since 2000. The `savings'
ingredient in the ideal-gas model, required for getting the Gamma function form
of the otherwise ideal gas (Gibbs) distribution, was also discovered more than a
decade earlier by John Angle. He employed a different driver
in his stochastic model of inequality process. This inequality 
coming mainly from the stochasticity, together with the equivalent 
of saving introduced in the model.
A proper Pareto tail of the Gamma distribution comes naturally in this
class of models when the saving propensity of the agents are distributed,
as noted first by the Kolkata group and analyzed by them and
by the Dublin group led by Peter Richmond.

Apart from the intensive involvements of physicists together with 
a few economists in
this new phase of development, a happy feature has been that
econophysics has almost established itself as a (popular) 
research discipline in statistical physics. 
Many physics journals have started publishing papers in such an
interdisciplinary field. Also, courses in econophysics 
are being offerred in several universities, mostly in their physics departments.

In spite of all these, it must be stated that there has, so far, been no
spectacular success. Indeed, the mainstream economists are yet to take note of
these developments. In her account, reporting on the Econophys-Kolkata I
(New Scientist, UK, 12 March 2005 issue, pp.6-7), Jenny Hogan reported
several criticisms by economists, mostly appreciating the observations, but
not the modelling efforts! The same kind of criticism have recently been 
expressed more emphatically by economists Mauro Gallegati, Thomas Lux and others
(http://www.unifr.ch/econophysics; doc/0601001; Physica A in press). 
We have included a few responses by physicists
like Peter Richmond and social statistician
like John Angle in this volume. Economist J Barkley Rosser, in his intriguing 
reflections (in the following paper) on these new developments of econophysics,
detects the same stories of the past becoming present, predicting the future
of econophysics to become a past story of economics soon!

\subsubsection*{Acknowledgments}
I am grateful to Anindya-Sundar Chakrabarti and Arnab Chatterjee for careful
reading of this manuscript and useful criticisms.

\end{document}